\documentclass[twocolumn]{aastex63}
\usepackage{rotating}
\usepackage{graphicx}
\usepackage{amssymb}
\usepackage{amsmath}
\usepackage{hyperref}

\def\gtrsim{\mathrel{\hbox{\rlap{\hbox{\lower4pt\hbox{$\sim$}}}\hbox{$>$}}}}
\def\lesssim{\mathrel{\hbox{\rlap{\hbox{\lower4pt\hbox{$\sim$}}}\hbox{$<$}}}}
\def\gtrsim{\mathrel{\hbox{\rlap{\hbox{\lower4pt\hbox{$\sim$}}}\hbox{$>$}}}}
\def\fdeg{\hbox{$.\!\!^{\circ}$}}
\def\farcs{\hbox{$.\!\!^{\prime\prime}$}}

\begin{document}

\title{{Chandra}, { NuSTAR}, and Optical Observations of the Cataclysmic Variables IGR~J17528$-$2022 and IGR~J20063+3641 }
\author{Jeremy Hare}
\altaffiliation{NASA Postdoctoral Program Fellow}
\affiliation{NASA Goddard Space Flight Center, Greenbelt,  MD 20771, USA}
\author{Jules P. Halpern}
\affiliation{Department  of Astronomy, Columbia  University, 550  West  120th Street, New  York, NY 10027, USA}
\author{John  A. Tomsick}
\affiliation{Space Sciences Laboratory, 7 Gauss Way, University of California, Berkeley, CA 94720, USA}
\author{John  R. Thorstensen}
\affiliation{Department of Physics and Astronomy, Dartmouth  College, Hanover NH  03755, USA}
\author{Arash Bodaghee}
\affiliation{Department of Chemistry, Physics and Astronomy, Georgia College and State University, Milledgeville, GA 31061, USA}
\author{Ma{\"\i}ca Clavel}
\affiliation{Univ. Grenoble Alpes, CNRS, IPAG, F-38000 Grenoble, France}
\author{Roman Krivonos}
\affiliation{Space Research Institute of  the Russian Academy of Sciences (IKI), Moscow, Russia, 117997}
\author{Kaya Mori}
\affiliation{Columbia Astrophysics Laboratory, Columbia University, New York, NY 10027, USA}

\email{jeremy.hare@nasa.gov}

\begin{abstract}
We report on {Chandra}, {NuSTAR}, and MDM observations of two {INTEGRAL} sources, namely IGR~J17528$-$2022 and IGR~J20063+3641. IGR~J17528$-$2022 is an unidentified {INTEGRAL} source, while IGR~J20063+3641 was recently identified as a magnetic cataclysmic variable (mCV)  by \cite{2018AJ....155..247H}. The {Chandra} observation of IGR~J17528$-$2022 has allowed us to locate the optical counterpart to the source and to obtain its optical spectrum, which shows a strong H$\alpha$ emission line. The optical spectrum and flickering observed in  the optical time-series photometry in combination with the X-ray spectrum, which is well fit by an absorbed partially covered thermal bremsstrahlung model, suggests that this source is a  strong mCV candidate. The X-ray observations of IGR~J20063+3641 reveal a clear modulation with a period of 172.46$\pm0.01$ s, which we attribute to the white dwarf spin period. Additional MDM spectroscopy of the source has also allowed for a clear determination of the orbital period at 0.731$\pm0.015$ d. The X-ray spectrum of this source is also well fit by  an absorbed partially covered thermal bremsstrahlung model. The X-ray spectrum, spin periodicity, and orbital periodicity allow this source to be further classified as an intermediate polar.
\end{abstract}

\section{Introduction}
The { International Gamma-ray Astrophysics Laboratory} ({INTEGRAL}; \citealt{2003A&A...411L...1W}) has played a pivotal role in uncovering the relatively faint end ($F_X\gtrsim10^{-11}$ erg cm$^{-2}$ s$^{-1}$) of the Galactic hard X-ray source population (see e.g., \citealt{2016ApJS..223...15B,2016MNRAS.460..513T,2017MNRAS.470..512K,2020arXiv200810665L}). While {INTEGRAL} has uncovered many new sources, a large fraction of them still remain unidentified due to their large positional uncertainties ($\sim 1'-4'$), 
 making it difficult to confidently locate their longer wavelength counterparts. Therefore, we have been carrying out a { NuSTAR} legacy survey to uncover the nature of a number of  hard Galactic X-ray sources discovered by {INTEGRAL} (see e.g., \citealt{2016MNRAS.461..304C,2019ApJ...878...15H,2019ApJ...887...32C}; for recent results). { NuSTAR}'s hard X-ray sensitivity allows one to more reliably characterize the unidentified source's hard X-ray spectrum. { Chandra} observations have also been obtained for several sources and, with its superb angular resolution, allows one to confidently locate the hard X-ray source's optical/NIR counterpart, which then can be followed up with optical/NIR spectroscopy. 

Galactic { INTEGRAL} sources belong to  a variety of different source classes, including isolated pulsars and their wind nebulae, high and low mass X-ray binaries hosting either a black hole or neutron star, and supernova remnants. However, among the most numerous Galactic source types uncovered by { INTEGRAL} are cataclysmic variables (CVs; see e.g., \citealt{2016ApJS..223...15B}),  which consist of a white dwarf (WD) that accretes material via Roche-lobe overflow from a late-type main sequence companion. CVs typically have orbital periods ($P_{\rm orb}$) of $\sim1-10$ hours \citep{2017PASP..129f2001M} and are often characterized as  being either magnetic or non-magnetic based on the magnetic field strength of the WD (i.e., having $B_{\rm WD}\gtrsim1$ MG or $B_{\rm WD}\lesssim1$ MG for mCVs or non-magnetic CVs,  respectively). The mCVs, which are most commonly detected at hard X-ray energies (see e.g., \citealt{2006MNRAS.372..224B,2017PASP..129f2001M,2020AdSpR..66.1209D}), can be further subdivided into polars and intermediate polars (IPs) depending on the strength of the WD's magnetic field, having $B_{\rm WD}\approx10-230$ MG or $B_{\rm WD}\approx1-10$ MG, for polars and IPs, respectively  (see e.g.,  \citealt{1990ApJ...350L..13C,1997A&A...327..183B,2015SSRv..191..111F}),  and its effects on the accretion flow. In polars, the WD has a strong enough magnetic field to channel a large fraction of the accretion flow onto its magnetic poles, thus preventing the formation of an accretion disk. As a result of the strong magnetic field of the WD locking on to the companion, the WDs in these systems are typically found to have spin periods ($P_{\rm spin}$) equal to the orbital period of the system. On the other hand, IPs contain WDs with intermediate strength magnetic fields, which allow for the formation of an accretion disk that is truncated by the WD's magnetosphere from where the accreted material is channeled onto the WD's magnetic poles. These systems typically have WD spin periods of 10 s to a few 1000 s with $P_{\rm spin}<P_{\rm orb}$ (see e.g., Figure 4 in \citealt{2020AdSpR..66.1209D}).

The hard X-ray emission from mCVs is thought to be produced by thermal bremsstrahlung emission from the shocked material in the accretion column above the surface of the WD.  This material exhibits a multitude  of temperatures, but typically has a peak temperature of tens of keV (see e.g., \citealt{2015ApJ...807L..30M,2017MNRAS.470.4815B}). X-ray emission from this multi-temperature thermal plasma can also excite neutral and ionized Fe on the WD surface/pre-shock accretion flow or in the post-shock accretion flow for neutral and ionized Fe, respectively,\ leading to strong Fe line features at 6.4 keV (neutral), 6.7 keV (He-like), and 6.97 keV (H-like; see e.g., \citealt{1999ApJS..120..277E,2001MNRAS.327..208W,2004MNRAS.352.1037H}).

IGR J17528$-$2022 and IGR J20063+3641 (J17528 and J20063 hereafter, respectively) are two { INTEGRAL} sources that were part of our {  NuSTAR} legacy survey.  J17528 was first discovered by the {Swift}-BAT at the $\sim12\sigma$ level and was reported in the 4th Palermo  { Swift-BAT} catalog (i.e., 4PBC J1752.6-2020; \citealt{2014styd.confE.132C}). The source was subsequently detected by { INTEGRAL} at the $6.7 \sigma$ level and reported in the catalog of \cite{2017MNRAS.470..512K}. In both catalogs, the source was designated as unidentified, but was observed and detected at soft X-ray energies by { Swift-XRT}. However, the positional accuracy of { Swift}-XRT did not afford a unique optical/NIR counterpart to the source, so it remained unclassified. 

J20063 was reported in the 70 month { Swift}-BAT catalog, having a detection significance of $\sim8.7\sigma$ (i.e., Swift J2006.4+3645; \citealt{2013ApJS..207...19B,2018ApJS..235....4O}) and a soft X-ray { Swift}-XRT counterpart. \cite{2017MNRAS.470..512K} also reported the detection of this source by { INTEGRAL} at the $9.1 \sigma$ level. During the time of the { NuSTAR} and { Chandra} observing campaign reported here, \cite{2018AJ....155..247H} identified the optical counterpart of the XRT source and obtained spectra that showed Balmer emission lines, and a \ion{He}{2} $\lambda4686$ line that had a comparable strength to the H$\beta$ line. They classified J20063 as a nova-like variable, or possibly a magnetic CV, at a distance of $\sim1-4$ kpc.  A lower-limit of 0.25~d was placed on the spectroscopic period of the system, with the strongest candidates at 0.421~d and 0.733~d.  Time-series photometry showed a peak in the power spectrum at 172~s, which the authors (mistakenly, it turns out; see Section \ref{results}) attributed to a multiple of their 43~s sampling period.

Here we report on { NuSTAR} and { Chandra} observations of J17528 and J20063. The precise X-ray localization of J17528 has also allowed us to identify the source's optical counterpart and to obtain its optical spectrum and optical time-series photometry, which is also presented here.  Additionally,  new optical spectra of J20063 resolve the ambiguity of its orbital period.  The paper layout is as follows: in Section \ref{oadr} we discuss the observations and data reduction, in Section \ref{results} we discuss our timing and spectral analyses, and in Section \ref{discuss} we discuss  where these sources lie in the broader X-ray emitting cataclysmic variable population (i.e., polars versus IPs). Lastly, we summarize our findings in Section \ref{summary}.

\section{X-ray Observations and Data Reduction}
\label{oadr}
\subsection{Chandra}
J17528 and J20063 were observed with the { Chandra} Advanced CCD Imaging Spectrometer (ACIS; \citealt{2003SPIE.4851...28G}) on 2018 April 27 (MJD 58236.0; ObsID 20199) and 2018 February 25 (MJD 58174.9; ObsID 20198), respectively. Both sources were observed by the back-illuminated ACIS-S3 chip operated in timed exposure mode and the data were telemetered using the ``faint'' mode. The sources were observed using a 1/8 sub-array in order to reduce the frame time to 0.4 s so that the pile-up remained $<2\%$ throughout the observations. J17528 was observed for 4.6 ks, while J20063 was observed for 4.51 ks. The { Chandra} data analysis reported here was performed using the { Chandra} Interactive Analysis of Observations (CIAO) software version 4.11 and the 4.8.3 version of the Calibration Database (CALDB). Prior to analysis, both event files were reprocessed with the CIAO tool {\tt chandra\_repro}.

The CIAO task {\tt wavdetect} was run on the 0.5-8 keV band image to identify all sources detected by { Chandra} in these two observations. In the field of J17528 only one point source was significantly detected (i.e., $>3\sigma$) at the location, R.A.$=268.\!^{\circ}205423$ and decl.$=-20.\!^{\circ}404359$, having a statistical positional uncertainty of $0\farcs12$ (estimated using Equation 12 from \citealt{2007ApJS..169..401K}). Similarly, only one point source was significantly detected in the field of J20063 at the location R.A.$=301.\!^{\circ}593318$ and decl.$=+36.\!^{\circ}695425$,  with a statistical positional uncertainty of 0$\farcs12$ estimated in the same way as above. Unfortunately, since no additional sources are detected in either field, we are unable to correct for any systematic offset in the absolute astrometry. Therefore, we account for this uncertainty by adopting { Chandra's} 90\% overall astrometric uncertainty of $0\farcs8\footnote{\url{http://cxc.harvard.edu/cal/ASPECT/celmon/}}$, which we convert to the 95\% uncertainty by multiplying by 2.0/1.7. We then add the statistical and systematic positional uncertainties together in quadrature and find a $2\sigma$ positional uncertainty of $0\farcs95$ for each source. 

The { Chandra} energy spectra and barycentered event arrival times (corrected to the solar system barycenter using the CIAO tool {\tt axbary} prior to extraction) for these sources were extracted from a 2$''$ radius circular region\footnote{On-axis a 2$''$ radius circle encloses $\sim$95\% of the {\sl Chandra} PSF at 1.5 keV, see Figure 4.6 here: \url{http://cxc.harvard.edu/proposer/POG/html}} centered on each source's {\tt wavdetect} position. The background spectra were extracted from source free annuli ($5''<r<20''$) also centered on each source's position. The energy spectrum of J17528 contained 316 net counts, while the energy spectrum of J20063 contained 395 net counts. We binned both spectra to have a signal-to-noise ratio of at least  five per energy bin. Unless otherwise noted, all uncertainties in this paper are reported at the 1$\sigma$ level.

\subsection{NuSTAR}
\label{nu_analy}
J17528 was observed by the { Nuclear Spectroscopic Telescope Array} ({ NuSTAR}; \citealt{2013ApJ...770..103H}) on 2018 May 9 (MJD 58247.6; ObsID 30401004002) for $\sim$ 43 ks, while J20063 was observed by { NuSTAR} on 2018 March 23 (MJD 58200.1; ObsID 30401003002) for $\sim36$ ks. Both datasets were reduced using the { NuSTAR} Data Analysis Software (NuSTARDAS) version 1.8.0 with CALDB version 20190410. The datasets were also both filtered for the increase in background flares caused by { NuSTAR}'s passage through the South Atlantic Anomaly by using the options {\tt saacalc=2}, {\tt saamode=optimized}, for both sources, and with {\tt tentacle=no} for J17528 and {\tt tentacle=yes} for J20063, reducing the exposure times to $\sim42$ ks and $\sim35$ ks, respectively. 

\begin{figure*}
\centering
\includegraphics[trim={0 0 0 0},width=18.0cm]{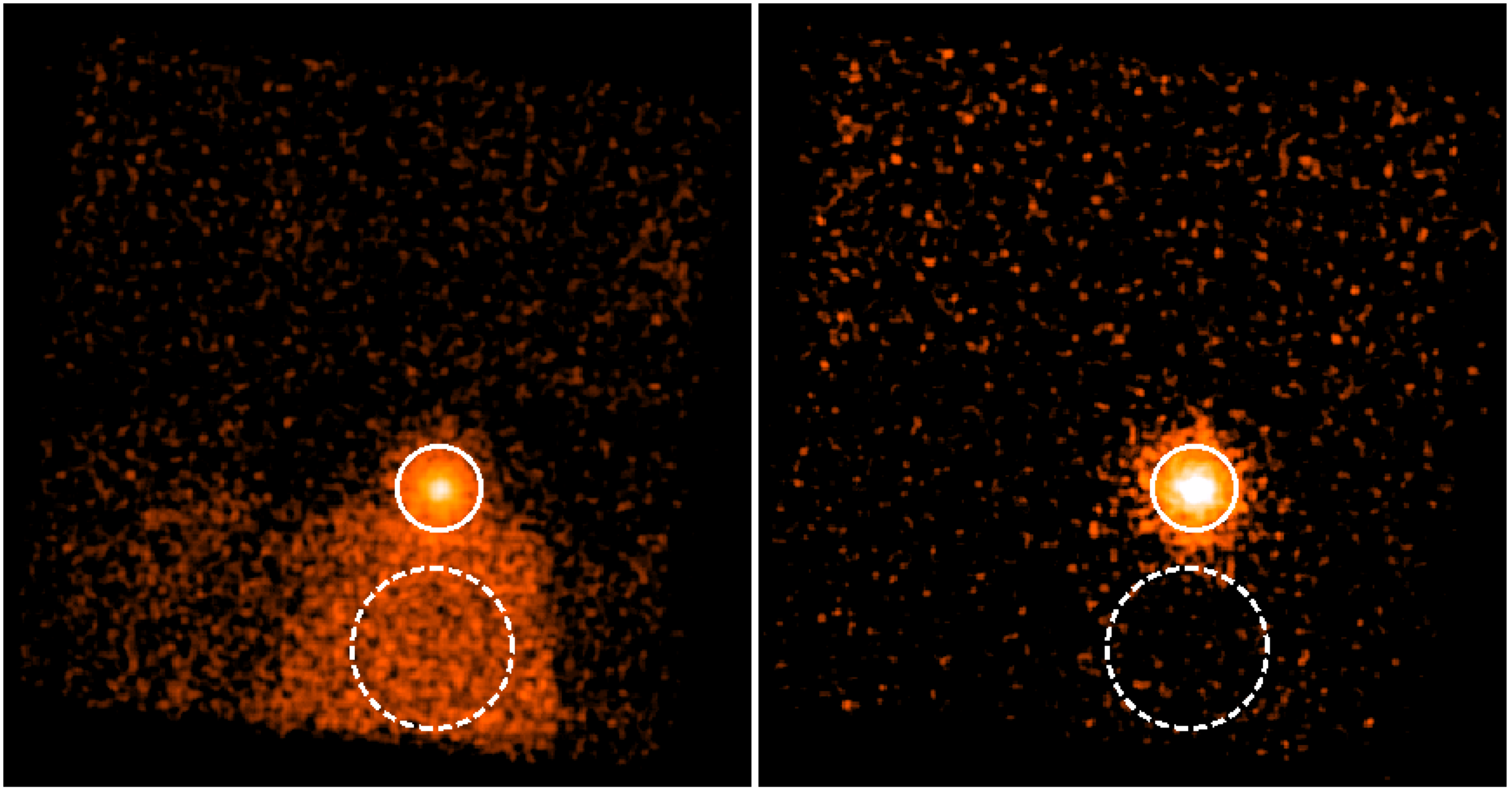}
\caption{{\sl Left:} { NuSTAR} FPMA 3-79 keV image of J20063. Strong absorbed stray light (likely from Cyg X-1; see Section \ref{nu_analy}) overlaps the source. The solid and dashed white circles show the source and background extraction regions for the source's energy spectra and light curves. {\sl Right:} Same image as on the left but filtered to contain only the 3-20 keV energy range. The strong absorbed stray light is effectively filtered out using this energy cut. 
\label{stray}
}
\end{figure*}

The { NuSTAR} energy spectra and barycentered event arrival times (corrected to the solar system barycenter using the {\tt barycorr} tool) for each source were extracted from a circular aperture ($r=70''$ and $r=50''$, for J17528 and J20063, respectively) centered on the source. The background energy spectra for J17528 were extracted from a source free circular region ($r\approx2'$) placed on the same detector chip as the source. The observation of J20063 suffered from strong absorbed stray light in the FPMA detector, likely from Cygnus X-1 which is located $\sim2\fdeg2$ away from J20063. Fortunately, this absorbed stray light is only strongly observed above 20 keV (see Figure \ref{stray} and Section 4.2 in \citealt{2017JATIS...3d4003M} for more details). Additionally,  the FPMB detector also suffered from normal stray light. To account for the absorbed and standard stray light, we extracted the background energy spectra for J20063 from a circular region ($r\approx1.6'$) placed on the regions containing the stray light, but still on the same detector chip as the source (see Figure \ref{stray}). Due to the absorbed stray light, and the fact that the source becomes background dominated above $\sim20$ keV, we limit our { NuSTAR} analysis of J20063 to the $3-20$ keV energy range. The { NuSTAR} spectra for both sources were grouped to have a signal-to-noise ratio of at least five per energy bin. J17528 has no spectral energy bins with a signal-to-noise ratio $>$5 above 30 keV, so the spectrum does not extend beyond this energy (note that the the FPMB spectrum only extends up to $\sim 25$ keV).

\subsection{ Swift-BAT}
\label{SBAT_analy}

 We use Swift-BAT data to  extend the energy range coverage for J20063  due to the  absorbed stray light in NuSTAR  (see Section \ref{nu_analy}). The spectrum, covering the 14-195 keV energy range, were taken from the Swift-BAT 105-Month Hard X-ray Survey\footnote{\url{https://swift.gsfc.nasa.gov/results/bs105mon/}} \citep{2018ApJS..235....4O}.  However, the source  becomes background dominated above $\sim100$  keV so we limit  our Swift-BAT analysis to the 14-100 keV energy range. 

\section{Results}
\label{results}

\subsection{X-ray Timing}
The barycenter corrected event lists from both { Chandra} and { NuSTAR} were used to search for orbital and spin periodicity using the $Z^2_1$ test \citep{1983A&A...128..245B}. The false alarm probability (FAP) for the maximum $Z^2_{1}$  value found in each periodogram is calculated by multiplying $e^{(-Z^{2}_{\rm 1,max}/2)}$ by the number of trials, while the 1 $\sigma$ uncertainties are given by the frequency where the periodogram has decreased to $Z_{\rm 1,max}^{2}-1$. Additionally, to search for non-periodic variability, we constructed light curves with varying time bin sizes (i.e., 250~s, 500~s, 1~ks) for both the  Chandra and NuSTAR barycenter corrected event lists. We then fit a constant to these light curves to assess the significance of any variability. The light curves were constructed using the Stingray python package \citep{2019ApJ...881...39H} by removing 300~s from the beginning and end of each good time interval (GTI) to minimize possible effects from an increased background that may appear near the borders of GTIs (see e.g., Section 5 in \citealt{2015ApJ...800..109B}). The { Chandra} light curves were made using the $0.5-8$ keV energy band. For { NuSTAR}, we used the $3-30$ keV energy band for J17528, while for J20063 we only use the $3-20$ keV band due to the absorbed and standard stray light dominating above 20 keV (see Section \ref{nu_analy}).

Neither the { Chandra} nor the { NuSTAR} light curves of J17528 displayed any significant variability, with the highest variability significance being detected in the 500~s binned $3-30$ keV {\sl NuSTAR} light curves at the $\sim2.3\sigma$ level. The $Z^{2}_1$ test was run on the { Chandra} event list in the frequency range between $\nu\approx3.9\times10^{-4}-0.16$ Hz using $\sim16000$ equally spaced frequencies. The maximum $Z^{2}_1=19.92$ at a frequency of 0.142 Hz (7.03 s) corresponds to a false alarm probability of 74\%. The $Z^2_1$ test was also used to search the $3-30$ keV { NuSTAR} event list in the frequency range $\nu\approx4.7\times10^{-5}-1$ Hz over $\sim10^{5}$ equally spaced frequencies. The largest $Z^2_1=29.04$ occurs at a frequency of 0.653 Hz ($P=1.53$ s) and has a FAP of $5\%$. Therefore, we conclude that no significant spin/orbital periodicity is detected in this source and that it is not variable on $\sim0.25-1$ ks timescales.

\begin{figure}
\centering
\includegraphics[trim={0 0 0 0},width=8.5cm]{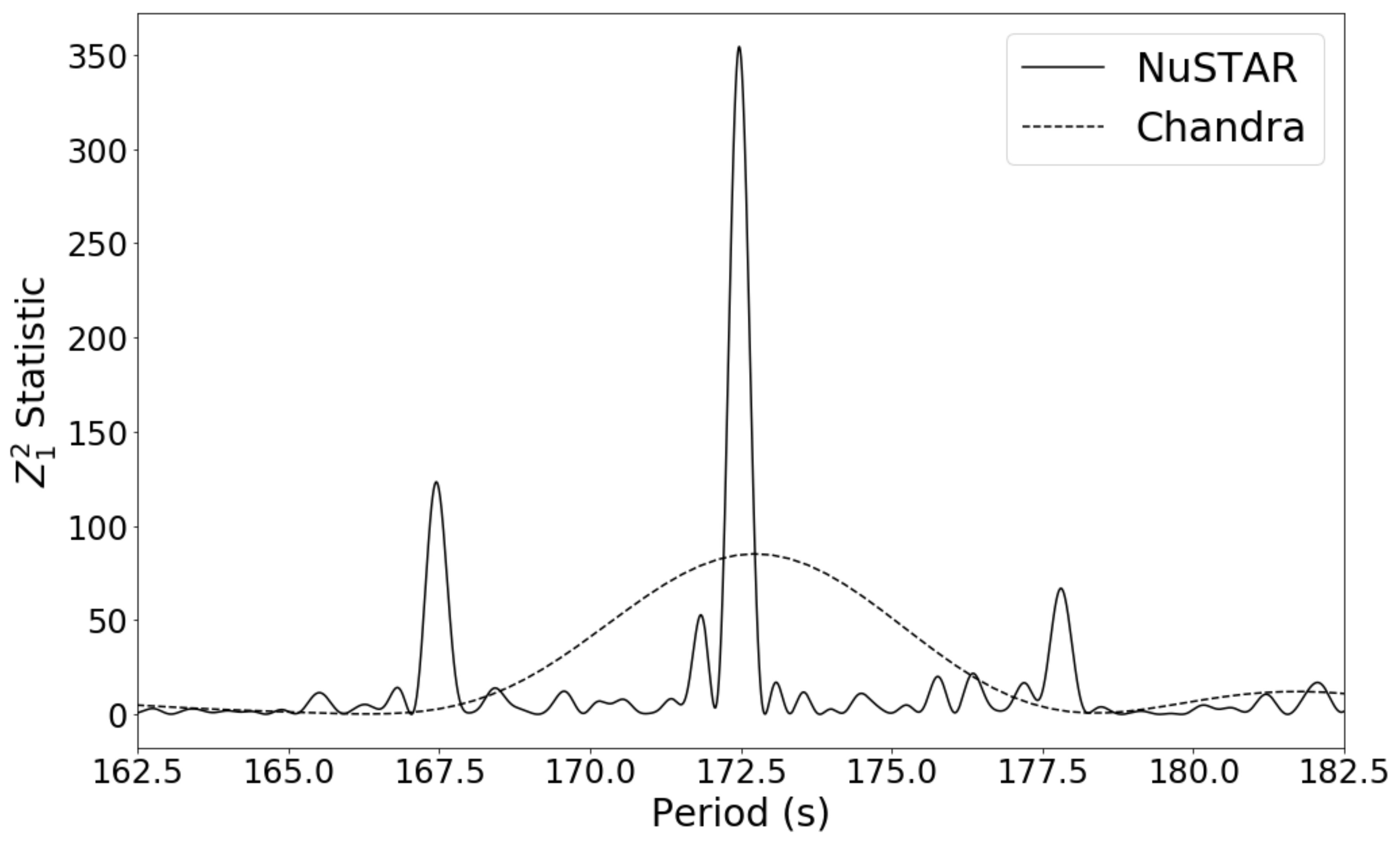}
\caption{Results of the $Z^2_1$ test for J20063. A strong periodic signal is detected in the { NuSTAR} and { Chandra} data of the source at a period $P_{\rm spin}\approx172.5$~s.
\label{zsq}
}
\end{figure}

J20063 also shows no indications of variability in the { Chandra} or { NuSTAR} light curves  on time scales $\geq250$ s. The highest variability significance was detected in the 500 s binned { NuSTAR} light curve at the 1.5 $\sigma$ level and the 250 s { Chandra} light curve at the $\sim2.1\sigma$ level. We also ran the $Z^{2}_{1}$ test on the { Chandra} event list over $\sim16000$ equally spaced frequency bins spanning the frequency range between $\nu\approx4.0\times10^{-4}-0.16$ Hz. Interestingly, a strong signal is detected in the { Chandra} periodogram at a frequency of $(5.79\pm0.01)\times10^{-3}$~Hz ($P=172.7\pm0.4$ s) with a maximum $Z^{2}_1=85.1$, corresponding to a FAP of $5\times10^{-15}$ (see Figure \ref{zsq}).  We attribute this short period to the spin of the WD.  We calculated the pulsed fraction of J20063 by taking the number of counts at the peak spin phase, subtracting off the number of counts at the minimum spin phase, and then divided by the sum of these two numbers. For { Chandra} the background contribution is negligible, and we find a pulsed fraction of $60\pm8\%$ in the $0.5-8$ keV energy range.

\begin{figure}
\centering
\includegraphics[width=8.5cm]{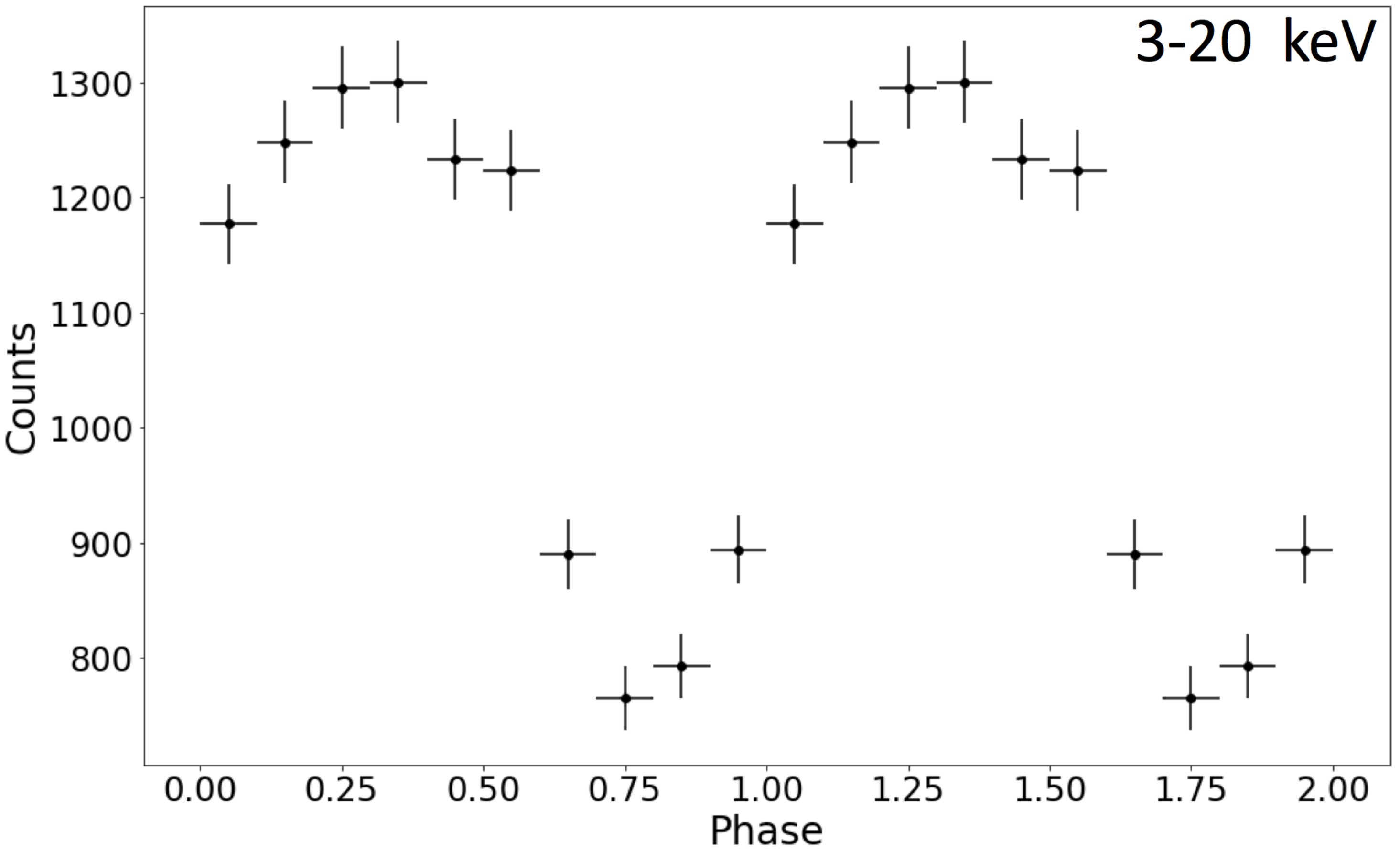}
\caption{{ NuSTAR} FPMA+FPMB pulse profile, folded at the $P_{\rm spin}=172.46$~s period, of J20063 in the 3-20 keV energy range.
\label{j20063_phased}
}
\end{figure}

A strong $Z^{2}_{1}$ peak is also detected in the { NuSTAR} data at a frequency of $(5.7983\pm0.0004)\times10^{-3}$  Hz ($P_{\rm spin}=172.46\pm0.01$ s) with a maximum $Z^{2}_1=361.4$ (see Figure \ref{zsq}).  The $3-20$ keV { NuSTAR} FPMA+B pulse profile, folded on the 172.46~s period, is shown in Figure \ref{j20063_phased}. The background contribution in { NuSTAR} is significant. Therefore, we calculate the pulsed fraction in the same way as above but subtract the phase averaged background from the maximum and minimum number of counts.  We find a pulsed fraction of 54$\pm5\%$ in the $3-20$ keV energy band. We also calculate a pulsed fraction of $57\pm7\%$ in the $3-10$ keV band and $51\pm8\%$ in the $10-20$ keV band.

\subsection{X-ray Spectra}
\label{spec}

\subsubsection{Non-reflection fits}
For the spectral analyses of both IGR sources, we simultaneously fit the { Chandra} and { NuSTAR} spectra in the $0.5-8$ keV and $3-30$ keV energy ranges for J17528, and $0.5-8$ keV and $3-20$ keV energy ranges for J20063, respectively. For J20063, we also simultaneously fit the 14-100 keV Swift-BAT data. In all fits discussed below, a multiplicative constant is used to account for possible calibration differences between the instruments. All energy spectra in this paper were fit using XSPEC version 12.10.1 \citep{1996ASPC..101...17A}. Additionally, we used the Tuebingen-Boulder ISM absorption model ({\tt tbabs}) with the solar abundances of \cite{2000ApJ...542..914W} in our fits.  We  note that the Galactic  absorbing  column densities are $N_{\rm H}=4.3\times10^{21}$ cm$^{-2}$ and $N_{\rm H}=9.7\times10^{21}$ cm$^{-2}$ in the directions of  J17528 and J20063, respectively \citep{2016A&A...594A.116H}.

The spectra for J17528 were first fitted with an absorbed power-law model.  The fitted model has a hard photon index of $\Gamma=0.91\pm0.03$ and relatively large absorbing column density, $N_{\rm H}=4.0^{+0.9}_{-0.7}\times10^{22}$ cm$^{-2}$. However, the quality of the fit is poor ($\chi^{2}/\nu=393/252$), showing large residuals around the Fe line complex at 6.4 keV and evidence for a spectral cutoff above $\sim20$ keV.  Therefore, we switch the continuum model to an absorbed thermal bremsstrahlung model (i.e., {\tt bremss} in Xspec; \citealt{1975ApJ...199..299K}), but the  fitted model has an unconstrained temperature, large residuals at soft X-ray energies, and provides a poor fit to the data ($\chi^{2}/\nu=450/257$). To overcome the large residuals at soft energies, we add partial covering absorber to the model. This model provides  the best fit to the continuum having a reduced chi-squared  $\chi^{2}/\nu=287/250$. Finally, we add a gaussian to the model to account for  the  neutral Fe K$\alpha$ line at 6.4 keV, leading to a reduced chi-squared  $\chi^{2}/\nu=252/248$ or, in other words, a $\Delta\chi^{2}=35$ for 2 fewer degrees of freedom. For clarity, the final model is {\tt const*tbabs*pcfabs*(bremss+gauss)}, which is hereafter referred to as Model 1. The best-fit parameters for  Model 1 are shown in Table \ref{tab_spec} while the residuals to this model are shown in Figure \ref{xrayspec_j17528}c.

J20063 was  previously identified as a CV \citep{2018AJ....155..247H}, so  we start with Model 1 (defined above) to fit its spectrum. Fitting the spectrum with a single gaussian leads to a large line width  ($\sigma_{\rm line}=0.49^{+0.13}_{-0.11}$ keV), which is larger than those typically observed in CVs (see e.g., \citealt{{2004MNRAS.352.1037H}}). To check for contributions from H-like and He-like Fe, we also fit a gaussian allowing the line center to be a free parameter. In this case, the line center shifts to $\approx$6.5 keV and the line width drops to $\approx350$ eV  (but is still consistent within the 1$\sigma$ uncertainty of the line width found when the gaussian was fixed at 6.4 keV).  The addition of a second gaussian does not statistically significantly improve the quality of the fit, nor does it significantly alter the neutral Fe line width. Thus, we use a single gaussian in the model for simplicity, as the data are not of high enough quality to constrain the contributions from ionized Fe lines, but note that there is likely some contribution from ionized Fe which is affecting the single-line parameters.

While Model 1 adequately fits the data, there are still systematic residuals at soft X-ray energies  (see Figure \ref{xrayspec_J20063}c). Therefore, we considered two additional models to account for these residuals. For the first model, we simply added an additional bremsstrahlung component to Model 1 (i.e., {\tt const*tbabs*pcfabs*(bremss+bremss+gauss)}; hereafter referred to as Model 2). This model was used as the emitting plasma in CVs often has a multi-temperature structure (see e.g., \citealt{2017PASP..129f2001M}). For the second model, we added a blackbody component to Model 1 (i.e., {\tt const*tbabs*pcfabs*(bremss+bbodyrad+gauss)}; hereafter referred to as Model  3) to account for these residuals.  These models further improve the fits by  $\Delta\chi^{2}=12$ and $11$, respectively for 2 fewer degrees of freedom and reduce the systematic residuals at low energies. An F-test gives a statistical significance of $\sim3\sigma$ for the additional bremsstrahlung or black body component.  For Model 2, we find that the additional bremsstrahlung component has a low temperature of 170$^{+40}_{-30}$ eV, which is similar to but larger  than the fitted  blackbody temperature of 122$^{+24}_{-21}$ eV. The best-fit parameters for  these models are shown in Table \ref{J200632_tab_spec} and the residuals for the fitted spectra are shown in Figure \ref{xrayspec_J20063}d,e. We have also added a second gaussian to these model but, similar to the non-blackbody model, found that it does not dramatically improve the quality of the fit nor does it significantly reduce  the Fe line widths, so we omit it.

\subsubsection{Reflection fits}

While the spectral models discussed above adequately fit the broadband X-ray spectra of both J17528 and J20063, a few of the best-fit parameters have somewhat extreme values. For instance, the partial covering absorption derived from these models is much higher than typically observed from IPs (i.e., the values are usually a few $\times10^{23}$ cm$^{-2}$; see e.g., \citealt{2007ApJ...663.1277E,2017MNRAS.470.4815B}). Additionally, the blackbody component in the spectral model of J20063 is on the hot end of the temperature distribution typically observed in IPs (see e.g., \citealt{2007ApJ...663.1277E,2008A&A...489.1243A,2017MNRAS.470.4815B,2020AdSpR..66.1209D}).  These issues motivated the use of a more complex spectral model that also accounts for the X-rays produced in the accretion column that are then reflected off of the WD's surface and have been observed in the X-ray spectra of several CVs (e.g., \citealt{2015ApJ...807L..30M}). To account for this component, we convolve the {\tt reflect} model \citep{1995MNRAS.273..837M} with the bremsstrahlung model.  The reflect model has several parameters\footnote{See \url{https://heasarc.gsfc.nasa.gov/xanadu/xspec/manual/node292.html}.}, including the reflection amplitude ($\Omega/2\pi$), abundances of elements heavier than He relative to solar ($A$), abundance of Fe relative to $A$ ($A_{\rm Fe}$), and the cosine of the inclination angle ($\cos{i}$). It has been noted that the partial covering model and reflection are degenerate with one another (see e.g., \citealt{2010A&A...520A..25Y,2016ApJ...826..160H}), thus including a reflection component can lower the hydrogen column density of the partial covering absorber.  Additionally, the reflection component can also lower the temperature of the plasma component. For clarity, the reflection model used for both sources is {\tt const*tbabs*pcfabs*(reflect*(bremss)+gauss)}, in XPSEC notation (hereafter referred to as Model  4), where the gaussian is again centered at 6.4 keV. We keep the gaussian component in this model because the  {\tt reflect} model does not account for fluorescent emission lines.  Note, for J20063, we exclude the blackbody component when fitting the reflection model.

\begin{figure}
\centering
\includegraphics[trim={20 0 0 0},width=8.5cm]{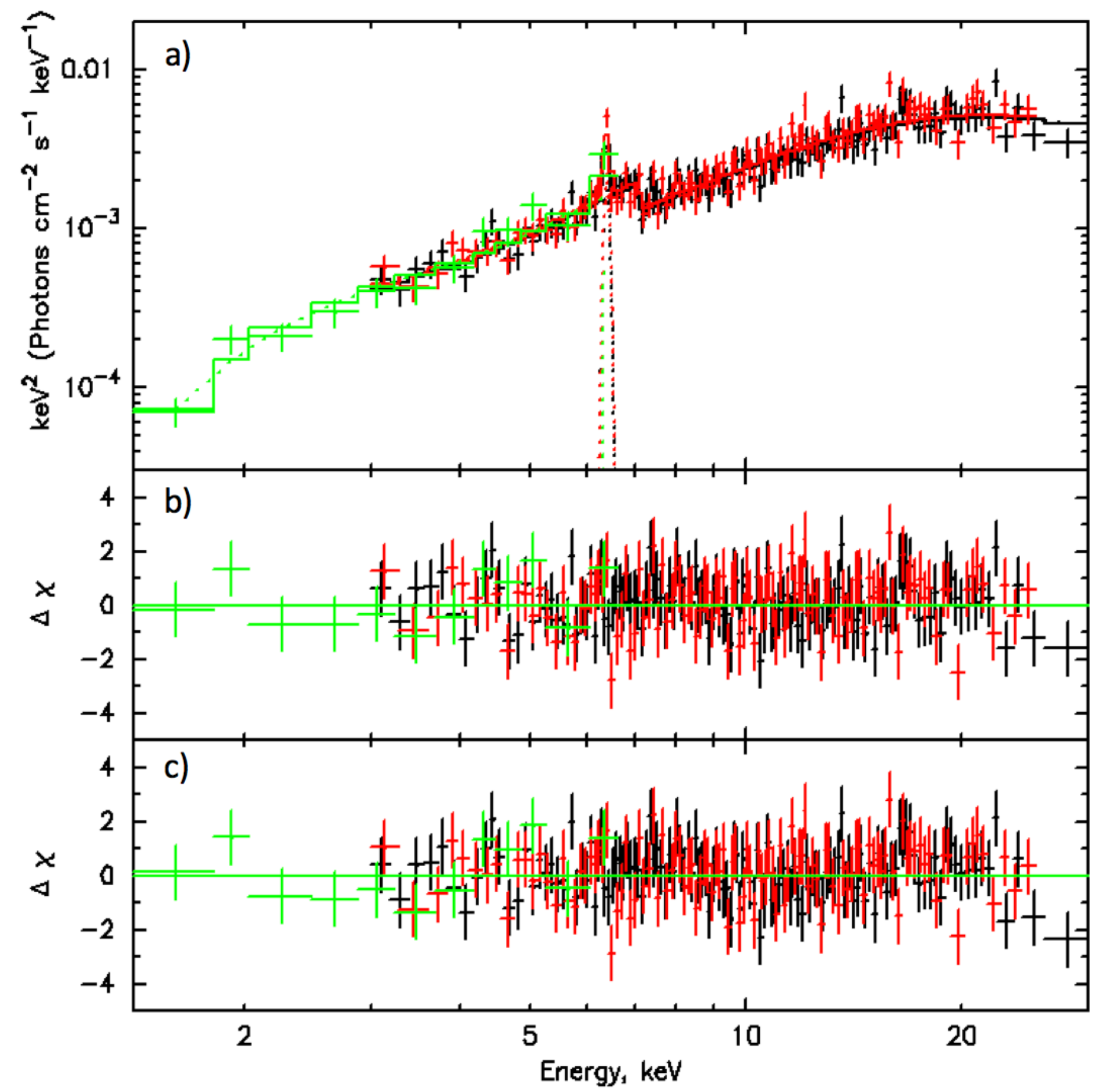}
\caption{{ Chandra} (green; $0.5-8$ keV) and { NuSTAR} FPMA (black; $3-30.0$ keV) and FPMB (red; $3\simeq25.0$ keV)  spectra for J17528 with the best-fit Model  4 (panel a) and residuals (panel b).  Panel c shows the residuals from the best-fit Model 1.
\label{xrayspec_j17528}
}
\end{figure}

While setting up the model, we explored leaving various parameters in the {\tt reflect} component free.  For both sources, we found that, if the reflection amplitude is left free, the fit prefers values $>1.0$. However, this value is unphysical under the condition that we see all of the direct emission (see e.g., \citealt{2016MNRAS.460..513T}). Therefore, we fix this parameter to a value of 1 in the fits for both sources. We also freeze the abundances, both A and $A_{\rm Fe}$, to 1 for both sources.
For J17528, Model 3 provides a better fit to the data (i.e., $\chi^{2}/\nu=241/247$ or $\Delta\chi^2$ of 11 for 1 fewer degree of freedom). The best-fit parameters provide a lower lower bremsstrahlung temperature (see Table \ref{tab_spec}). The best-fit reflection spectra and the residuals are shown in Figure  \ref{xrayspec_j17528}a,b. 

\begin{figure}
\centering
\includegraphics[trim={20 0 0 0},width=8.5cm]{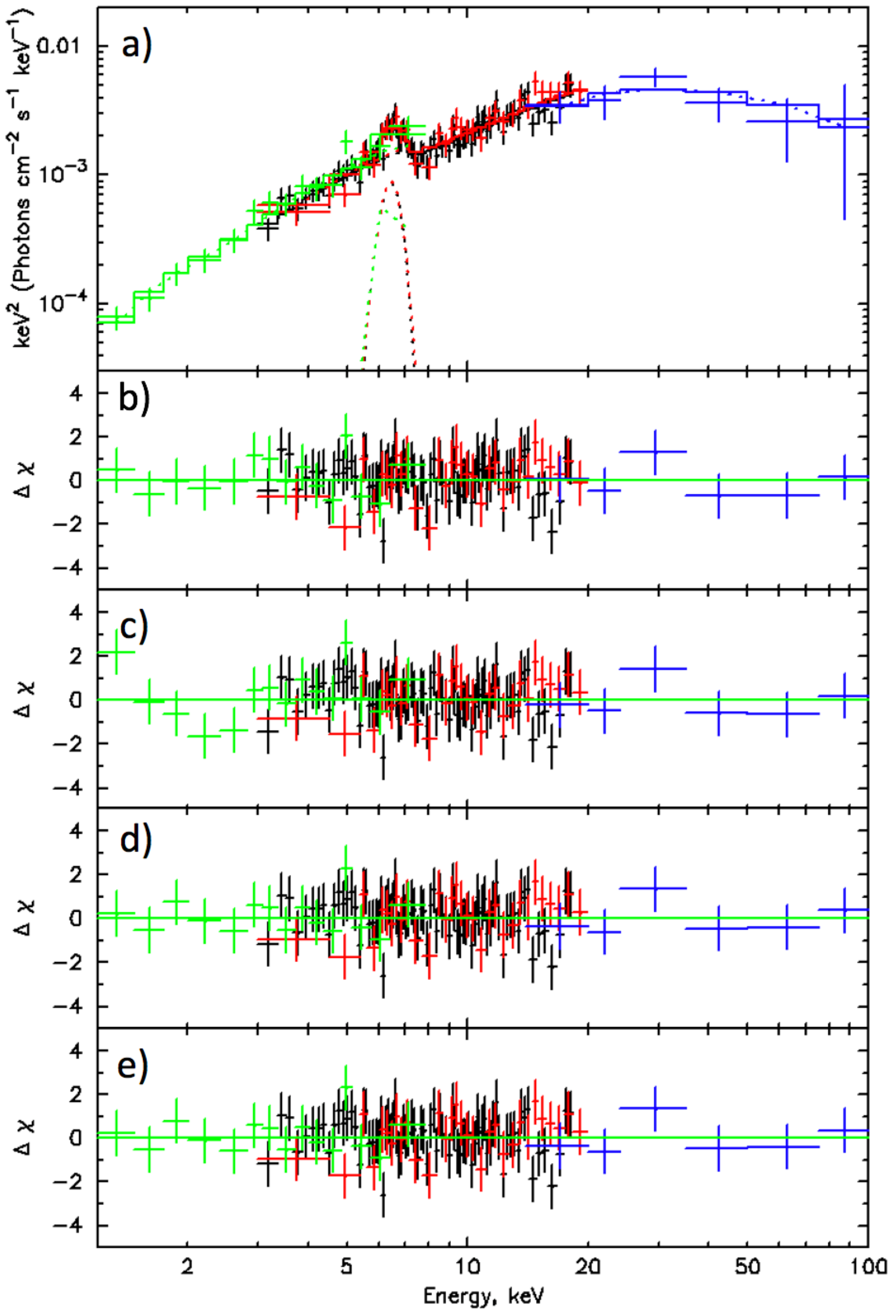}
\caption{{\bf a)} { Chandra} (green; $0.5-8$ keV), { NuSTAR} FPMA (black; $3-20.0$ keV), FPMB (red; $3-20.0$ keV),  and Swift-BAT (blue; $14-100$ keV) spectra for J20063 with the best-fit Model 4. {\bf b)} Residuals  for the best-fit Model  4. {\bf c)} Residuals for the best-fit Model 1. Systematic residuals are apparent in the $1-3$ keV energy range. {\bf d)} Residuals for the best-fit Model 2. {\bf e) Residuals for the best-fit Model 3.}
\label{xrayspec_J20063}
}
\end{figure}

For J20063, the reflection model provides a slightly worse fit to the data than Models 2 and 3, but has a more realistic value of the partial covering absorption  column density (see Table \ref{J200632_tab_spec}). Additionally, the reflection model eliminates the systematic residuals at soft X-ray energies without the need of a relatively  high temperature blackbody component, which  suggests that these residuals may be due to the lack of a reflection component in Model 1. Therefore, we favor the reflection models over the non-reflection models for both sources as they provide more reasonable physical values for the model parameters. The best-fit reflection spectral model for J20063 and the corresponding residuals are shown in Figure \ref{xrayspec_J20063}a,b.

\begin{table*}
\caption{Best-fit parameters derived from Model 1 and Model  4 for J17528 (see Section \ref{spec} for model details).} 
\label{tab_spec}
\begin{center}
\renewcommand{\tabcolsep}{0.11cm}
\begin{tabular}{lcccc}
\tableline 
Model Component & Parameter &  Unit &	Model 1 & Model 4 \\
\tableline
{\tt const} & FPMB/FPMA & -- & 1.04$\pm0.02$ &  1.04$\pm0.02$\\
& CXO/FPMA & -- & 1.02$\pm0.08$ &  1.04$\pm0.08$\\
\tableline 
{\tt tbabs} & $N_{\rm H}$ & $10^{22}$  cm$^{-2}$  & 3.4$\pm0.5$ &  3.2$\pm0.5$\\
\tableline 
{\tt pcfabs} & $N_{\rm H}$ & $10^{22}$  cm$^{-2}$ & 121$\pm15$ &  90$\pm14$\\
 & Covering Fraction & -- & 0.72$\pm0.03$ &  0.66$\pm0.03$\\
\tableline
{\tt gaussian} & $E$ &  keV & 6.4\tablenotemark{a} &  6.4\tablenotemark{a}\\ 
 & $\sigma$ &  keV & 0.10$^{+0.07}_{-0.09}$ &  0.049$^{+0.088}_{-0.046}$\\
 & $N$ &  10$^{-5}$ ph cm$^{-2}$ s$^{-1}$ & 2.1$^{+0.5}_{-0.4}$ &  1.4$^{+0.4}_{-0.3}$\\
 equivalent width &  & eV & 252$_{-50}^{+57}$ &  202$_{-43}^{+56}$ \\
\tableline 
{\tt reflect} &  $\Omega/2\pi$ & --  & -- & 1.0\tablenotemark{a}\\
 & $A$ & -- & --  &   1.0\tablenotemark{a}\\
&  $A _{\rm Fe}$ & -- &  -- &  1.0\tablenotemark{a}\\
 & $\cos{i}$ & -  &   --  &  $>$0.72\\
\tableline 
{\tt bremss} & $kT$ &  keV & 40$^{+13}_{-8}$ & 25$^{+4}_{-3}$\\
 & $N$ &  10$^{-3}$ & 1.8$\pm0.2$ & 1.3$\pm0.1$\\
 \tableline
Observed flux & $0.5-79.0$ keV & 10$^{-11}$ erg cm$^{-2}$ s$^{-1}$ &  1.65$^{+0.15}_{-0.12}$ &   1.36$\pm0.07$\\
 \tableline
$ \chi^2/$d.o.f. & & & 252/248 & 241/247\\
\tableline
\end{tabular} 
\end{center}
\tablenotetext{\textnormal{a}}{Fixed value.}
\end{table*}

\begin{table*}
\caption{Best-fit parameters derived from  Models 1, 2, 3, and 4 for J20063 (see Section \ref{spec} for model details).} 
\label{J200632_tab_spec}
\begin{center}
\renewcommand{\tabcolsep}{0.11cm}
\begin{tabular}{lcccccc}
\tableline 
Model Component & Parameter &  Unit & Model 1 & Model 2 & Model 3 & Model 4  \\
\tableline
{\tt const} & FPMB/FPMA & --  &  1.04$\pm0.04$ & 1.04$\pm0.04$ & 1.04$\pm0.04$ & 1.04$\pm0.04$  \\
& CXO/FPMA & -- & 1.06$\pm0.08$ & 1.14$^{+0.09}_{-0.08}$ & 1.14$^{+0.09}_{-0.08}$ & 1.15$^{+0.09}_{-0.08}$  \\ 
& BAT/FPMA & -- &  1.02$^{+0.19}_{-17}$ &  1.05$^{+0.19}_{-0.17}$ &  1.05$^{+0.19}_{-0.17}$ & 0.89$^{+0.17}_{-0.15}$  \\ 
\tableline 
{\tt tbabs} & $N_{\rm H}$ & $10^{22}$  cm$^{-2}$ &  2.6$\pm0.4$ & 5$\pm1$ & 5$\pm1$ &    1.26$^{+0.71}_{-0.69}$  \\ 
\tableline 
{\tt pcfabs} & $N_{\rm H}$ & $10^{22}$  cm$^{-2}$ &  131$^{+27}_{-26}$ & 163$^{+32}_{-29}$ &  162$^{+32}_{-29}$ &  18$^{+16}_{-6}$ \\
 & Covering Fraction & -- &  0.65$^{+0.04}_{-0.05}$ &  0.63$^{+0.05}_{-0.06}$ &    0.63$^{+0.05}_{-0.06}$ &  0.61$^{+0.13}_{-0.12}$  \\
\tableline
{\tt bbodyrad} & $kT$ & eV & -- & --&  122$^{+22}_{-20}$ & --  \\ 
 & Norm\tablenotemark{a} & 10$^{5}$ &  -- & --&  2.0$^{+17}_{-1.8}$ & -- \\
\tableline
{\tt gaussian} & $E$ &  keV & 6.4\tablenotemark{b} & 6.4\tablenotemark{b} &  6.4\tablenotemark{b} & 6.4\tablenotemark{b} \\  
 & $\sigma$ &  keV &  0.49$^{+0.13}_{-0.11}$ &  0.46$^{+0.11}_{-0.10}$ &    0.46$^{+0.11}_{-0.10}$ & 0.36$\pm0.09$  \\
 & $N$ &  10$^{-5}$ ph cm$^{-2}$ s$^{-1}$ &  5.3$^{+1.8}_{-1.4}$ &  5.5$^{+1.6}_{-1.3}$ &    5.5$^{+1.6}_{-1.3}$ & 2.2$\pm0.4$  \\
  equivalent width &  & eV &  760$_{-170}^{+160}$ & 720$^{+150}_{-160}$ &  720$_{-140}^{+150}$ &  560$\pm130$ \\
\tableline
{\tt reflect} &  $\Omega/2\pi$ & --  & -- & -- &  -- &1.0\tablenotemark{b} \\
 & $A$ & -- &  -- &-- &  --&1.0\tablenotemark{b} \\
&  $A _{\rm Fe}$ & -- &   -- & -- &  --&1.0\tablenotemark{b} \\
 & $\cos{i}$ & --  &    -- & -- &  --& $>$0.75 \\
\tableline  
{\tt bremss$_{1}$} & $kT$ &  keV &  42$^{+16}_{-10}$  &  37$^{+13}_{-9}$ & $ 37^{+13}_{-9}$ & $ 58^{+21}_{-15}$  \\
 & $N$ &  10$^{-3}$ &  1.54$^{+0.18}_{-0.14}$ & 1.7$\pm0.2$   &   1.7$\pm0.2$ &  0.77$^{+0.04}_{-0.03}$ \\
 {\tt bremss$_{2}$} & $kT$ &  keV &  --  &  0.17$^{+0.04}_{-0.03}$ & -- & -- \\
 & $N$ &  10$^{0}$ &  -- &  14$^{+93}_{-12}$   &  -- & -- \\
 \tableline
 Observed  flux & $0.5-100$ keV & 10$^{-11}$ erg cm$^{-2}$ s$^{-1}$ & 1.6$\pm0.2$ & 1.5$\pm0.1$ &   1.5$\pm0.2$ & 1.7$\pm0.2$ \\
 \tableline
$ \chi^2/$d.o.f. & & & 116/124 &  104/122 & 105/122 & 110/123 \\
\tableline
\end{tabular} 
\end{center}
\tablenotetext{\textnormal{a}}{Defined as $R^{2}_{\rm km}/D^{2}_{10}$, where $R_{\rm km}$ is the source radius in km and $D_{10}$ is the distance to the source in units of 10 kpc (see \url{https://heasarc.gsfc.nasa.gov/xanadu/xspec/manual/node139.html}.)}
\tablenotetext{\textnormal{b}}{Fixed value.}
\end{table*}

\subsection{Optical and Near Infrared Data}
\subsubsection{IGR J17528$-$2022}
\label{Optspec_17528}

\begin{figure}
\centering
\includegraphics[width=9.2cm, trim= 0 0 0 0]{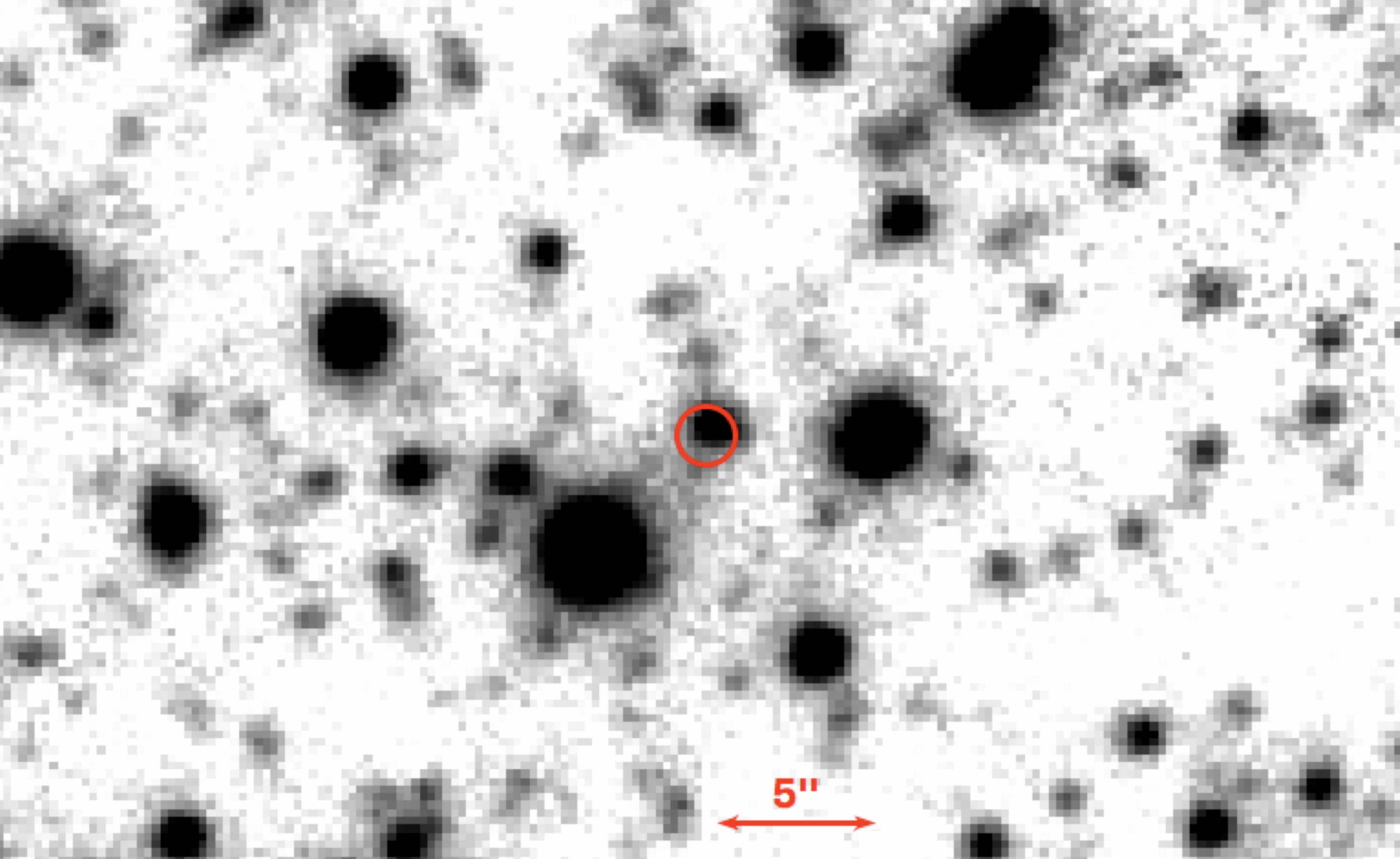}
\caption{{ Pan-STARRs y-band finding chart of the counterpart to J17528. The red circle, having a radius corresponding to the X-ray sources 2$\sigma$ positional uncertainty, shows the Chandra position of J17528. Up is north and east is to  the left. }
\label{find_chart}
}
\end{figure}

The { Chandra} observation of J17528 has allowed us to localize its position to an accuracy of about $\sim1''$, allowing us to identify its longer wavelength counterpart. Only one { Gaia} \citep{2020arXiv201201533G} optical counterpart is located within the positional error radius of the X-ray source.  The Pan-STARRS \citep{2016arXiv161205243F,2016arXiv161205560C} finding chart for this source is shown in Figure \ref{find_chart}. Unfortunately, the {\sl Gaia} counterpart has a negative parallax and a significant amount of astrometric excess noise ($\gtrsim10.0$; \citealt{2020arXiv201201533G}).  Therefore, its distance estimate of  $5.8$ kpc, inferred from the probabilistic method  of  \cite{2020arXiv201205220B},  is likely unreliable. 

\begin{figure}
\centering
\includegraphics[trim={0.cm 0 0 0},width=9.5cm]{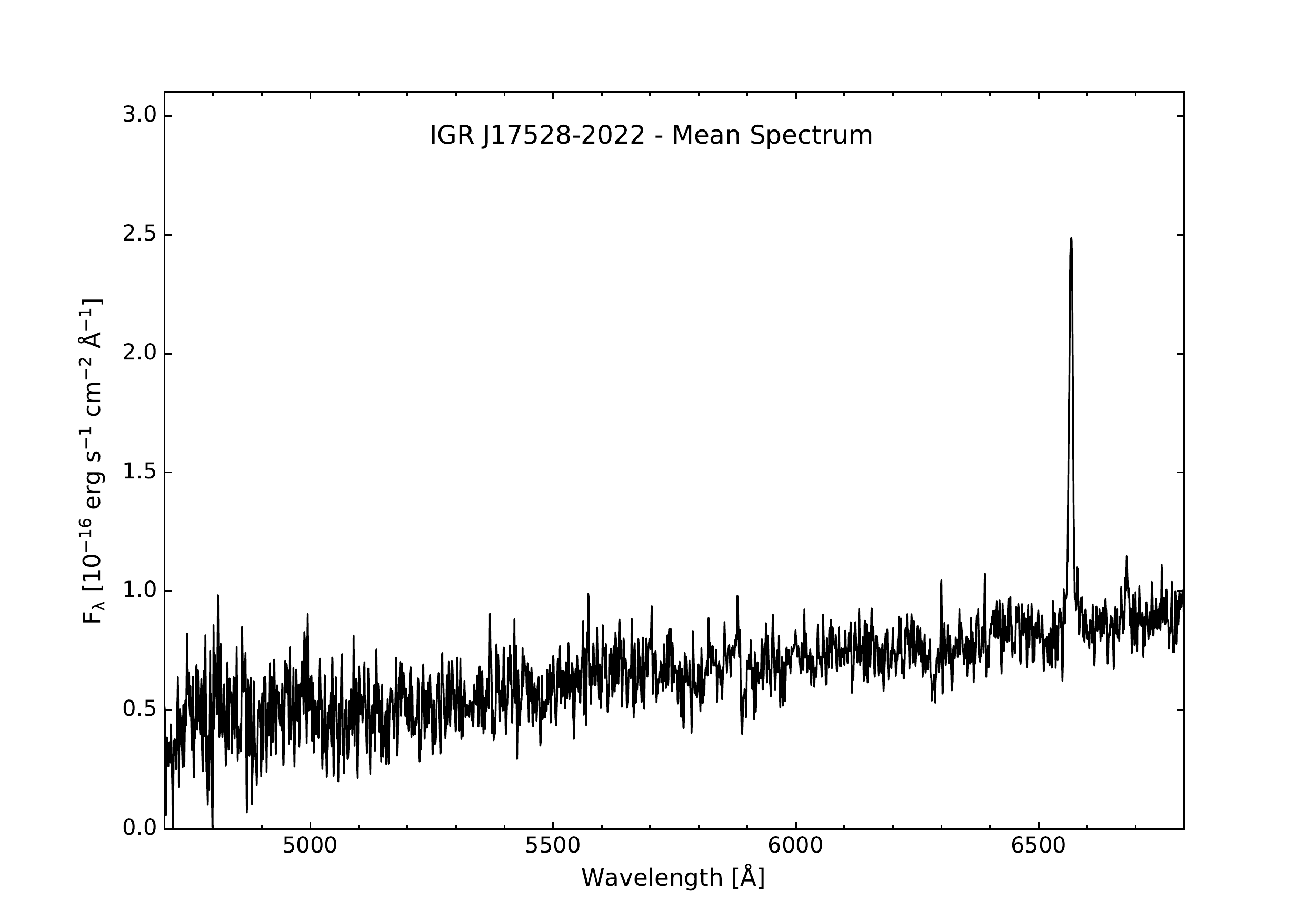}
\caption{{Mean MDM 2.4~m spectrum of J17528, from two 1000~s exposures.  The spectrum has been smoothed with a three-point running average, and the flux calibration is derived from observations of spectrophotometric standard stars.  H$\alpha\ \lambda6563$ emission is clearly visible, and there may be \ion{He}{1} $\lambda5876$ and $\lambda6678$ emission, as well as interstellar \ion{Na}{1}~D absorption and diffuse interstellar bands.}
\label{opt_spec}
}
\end{figure}

We observed J17528 on 2019 July 2 UT,
using the Ohio State Multi-Object Spectrograph
(OSMOS; \citealt{martini}) on the 2.4~m Hiltner
telescope at MDM Observatory on Kitt Peak, Arizona.
Two 1000-sec spectra were obtained, covering $4200-6800$~\AA\
at 0.7~\AA\ pixel$^{-1}$ and $\sim 3$~\AA\ resolution.
We also obtained three 20-s direct images through a Sloan $g$ 
filter immediately prior to the spectra, as part of 
the target acquisition.  We derived photometric 
zero points from the Pan-STARRS~1 (PS1) $g_{\rm PSF}$ magnitudes 
of stars in the field, and using these we find 
$g = 20.5 \pm 0.3$ for J17528, basically
identical to  its PS1 magnitude $g_{\rm PSF} = 20.48$ \citep{2016arXiv161205243F,2016arXiv161205560C}.

Figure~\ref{opt_spec} shows our mean spectrum.
The only clearly significant feature is emission
at H$\alpha$, with an equivalent width of $\sim 16$
\AA\ and a FWHM of $\sim 9$ \AA.  A hint of 
absorption may be seen near 6285 \AA\,  close to some
diffuse interstellar bands (DIBs; \citealt{jenniskens})
but also close to a telluric feature.  There also
may be some \ion{Na}{1}~D absorption ($\lambda 5889,\lambda5895$)
which, if present, would almost certainly be interstellar.
 The presence of the H$\alpha$ and He I lines suggest  that  the optical emission is likely coming from the WD accretion column or disk.

\begin{figure}
\centering
\includegraphics[width=9.2cm, trim= 17mm 0 0 0]{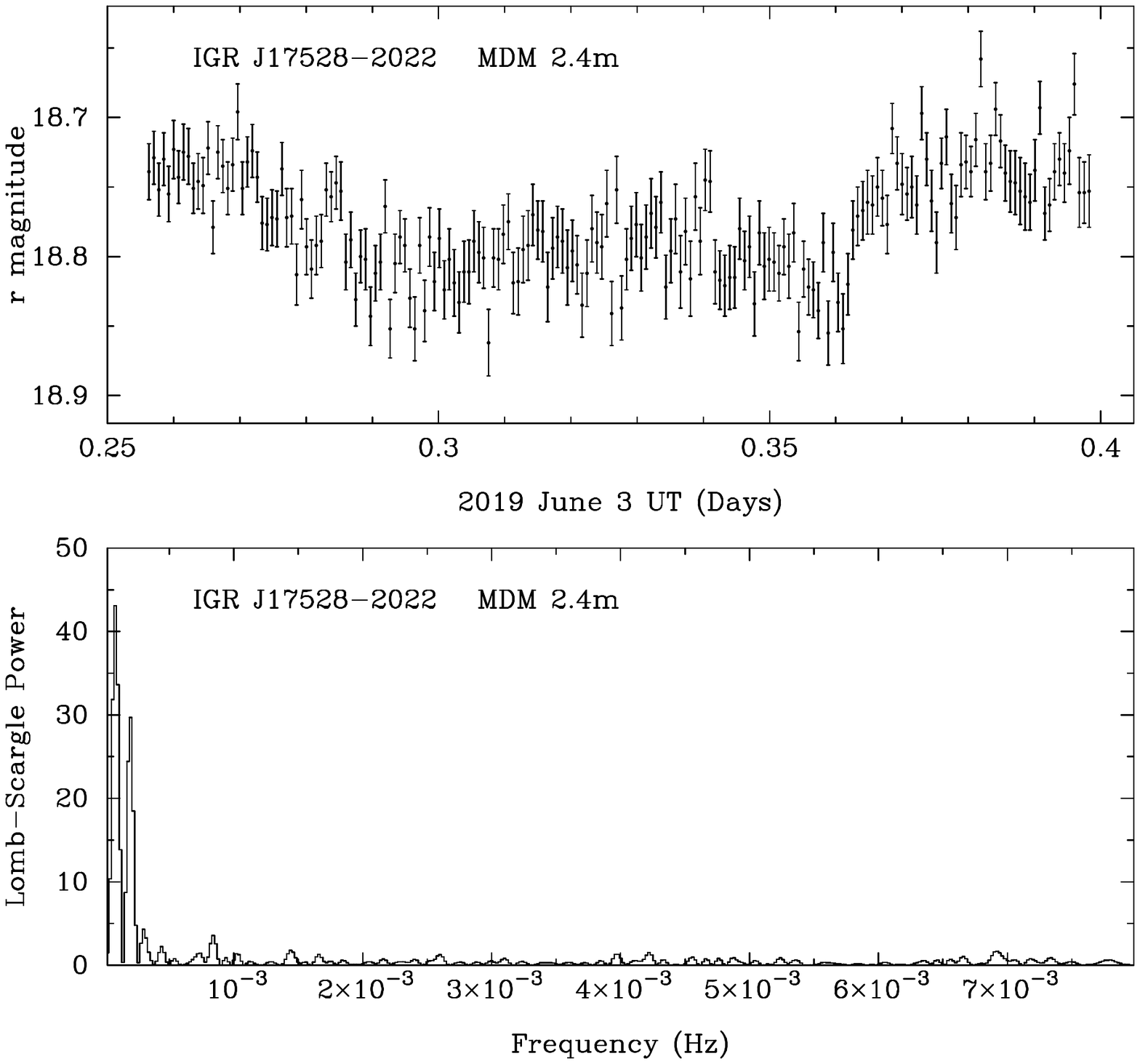}
\vspace{-4.7cm}
\caption{{Top: MDM 2.4~m $r$-band time-series photometry of J17528 at 64~s cadence.  The magnitude scale is referenced to Pan-STARRS photometry of a comparison star.  Bottom: Power spectrum of the time series. The light curve does exhibit variability, but no periodicity is detected, suggesting that any photometric period is $>3$~hr.}
\label{opt_phot}
}
\end{figure}

In addition to using the archival multiwavelength photometry, we obtained an $r-$band time series of the source on 2019 June 3, using OSMOS in direct imaging mode to perform differential photometry.  The duration of the observation was 3.4~hr at 64~s cadence (Figure \ref{opt_phot}).  While J17528 did exhibit some variability, no strong periodic signal was detected, so we are unable to further constrain the spin and/or orbital period of the source.

\subsubsection{IGR J20063+3641}

The optical counterpart of J20063 was first identified by \cite{2018AJ....155..247H} who suggested that the source is a nova-like variable, or a  mCV given that the \ion{He}{2} $\lambda4686$ and H$\beta$ lines were approximately the same strength.  A photometric signal at 172~s was detected, the same as the X-ray period found  here, but it was mistakenly attributed to a multiple of the 43~s sampling period. \cite{2018AJ....155..247H} could not find an unambiguous period from their  H$\alpha$ radial velocity  measurements, but did find candidate periods of $0.421\pm0.002$~d and $0.733\pm0.003$~d.  

 We obtained 15 more spectra on
2018 September 28 and 29, using the MDM Hiltner
telescope and modspec, configured as in
\cite{2018AJ....155..247H}; nine of these spectra gave
usable radial velocities of the H$\alpha$ emission.
On 2019 July 4 and 5 we obtained another 11 H$\alpha$
velocities with OSMOS (see Section \ref{Optspec_17528}). On these
nights the acquisition images (calibrated against
the Pan-STARRS $g$ magnitudes as described
earlier) showed the source at $g = 17.77$,
with little variation from night to night,
somewhat brighter than in Pan-STARRS (see below).
The top panel of Figure \ref{fig:igr2006montage}
shows the mean OSMOS spectrum.  The new velocities
resolve the period ambiguity firmly in favor of
the 0.73-day alias; the middle panel of
Figure \ref{fig:igr2006montage} shows the periodogram.
The number of cycles elapsed during the long intervals between observing runs is not determined, resulting in a cluster of fine-scale aliases and complicating the period uncertainty.  Examining fits at different aliases leads to $P_{\rm orb} = 0.731 \pm 0.015$~d as a conservative range.  The lower panel of Fig.~\ref{fig:igr2006montage} shows
the radial velocities folded on the single best-fitting
fine-scale period.

\cite{2018AJ....155..247H} also set an upper-limit on the distance to the source of $\sim4$~kpc by using the extinction map of \cite{2015ApJ...810...25G} and the typical unreddened absolute magnitudes of novalike variables. The optical counterpart is  detected by { Gaia}  \citep{2020arXiv201201533G}, which has measured its parallax,  $\pi=0.22\pm0.09$, and proper motion,  $\mu_{\alpha}\cos\delta=-4.30\pm0.09$ mas yr$^{-1}$, $\mu_{\delta}=-5.3\pm0.1$ mas yr$^{-1}$, with an insignificant  amount  of astrometric excess noise. The measured parallax has a large relative uncertainty   ($\sim40\%$), leading to  a large uncertainty in the inferred distance,  $d=4.5^{+1.4}_{-1.1}$ kpc \citep{2020arXiv201205220B}. The optical counterpart  of this source is also detected  by Pan-STARRS  \citep{2016arXiv161205243F,2016arXiv161205560C} at optical wavelengths, and by the UKIDSS survey \citep{2008MNRAS.391..136L} at NIR wavelengths. 

\begin{figure}
\centering
\includegraphics[width=9.2cm, trim= 17mm 0 0 0]{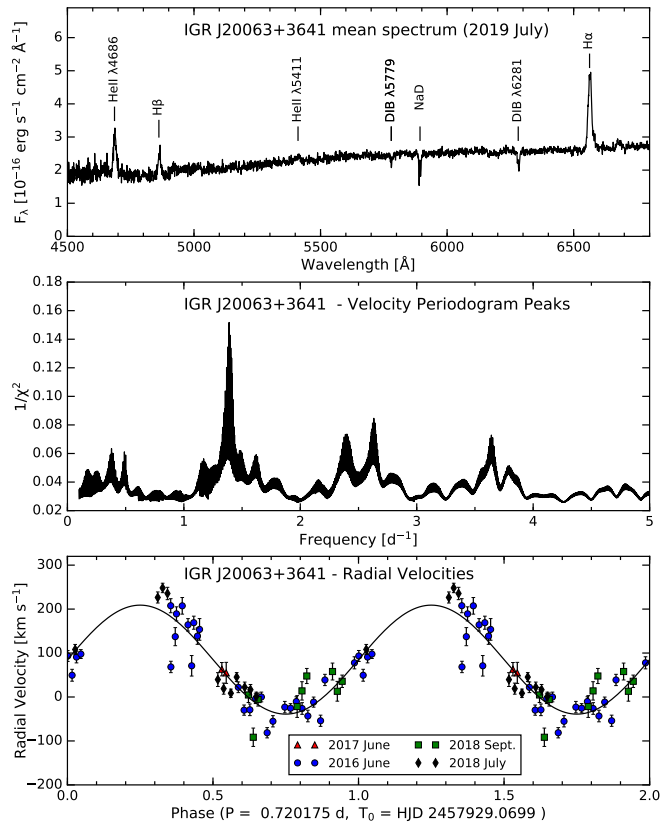}
\vspace{-1.0cm}
\caption{{Top:  Mean spectrum of IGR J20063+3641 from 2019 July.
Prominent emission lines and interstellar features are
marked.   Middle: Periodogram of the H$\alpha$ radial
velocities, formed by fitting a full sinusoid at each trial
frequency and inverting the squared residuals. Bottom: H$\alpha$ radial velocities folded on the single
best-fitting period; data are repeated over a second
cycle for continuity.  The 2018 September and 2019 July
data are newly reported, while the earlier data are
from \cite{2018AJ....155..247H}}
\label{fig:igr2006montage}
}
\end{figure}

\section{Discussion}
\label{discuss}
\subsection{IGR J17528$-$2022}
\label{17528_diss}

J17528's X-ray spectrum is well described by a partially covered thermal bremsstrahlung model with a narrow neutral Fe line and a possible reflection component. This type of spectrum is most typically observed in CVs (e.g., \citealt{2015ApJ...807L..30M,2016MNRAS.460..513T}).  Furthermore, the source's optical spectrum shows a strong H$\alpha$ emission line as well as weak He I lines, likely being produced by an accretion column or disk. The optical time-series photometry of J17528 shows flickering on minute-long time scales, which has also been observed in many CV systems (see e.g., \citealt{2018AJ....155..247H}). All of these factors strongly suggest that J17528 is a CV.

Since J17528's optical emission is likely dominated by the accretion column/disk, it is difficult to place constraints on the source's distance or the spectral type of the companion star using the multi-wavelength photometry. Converting the absorbing column density ($N_{\rm H}=3.2\times10^{22}$ cm$^{-2}$) from the best-fit X-ray spectrum to an optical absorption using the relation of \cite{2009MNRAS.400.2050G} provides $A_V=14.5$. The reddening map of \cite{2019ApJ...887...93G}  only  extends to $\sim2.5$ kpc in the direction of J17528 and provides $A_V=3.2$ at this distance. This suggests that J17528 is at a distance of at least a few kpc. Assuming a fiducial distance of  3 kpc to the source implies an observed X-ray luminosity of  $L_X\approx2\times10^{34}$ erg s$^{-1}$ in the 0.5-79 keV energy range.  Unfortunately, no spin or orbital period was detected in the optical or X-ray data, therefore, we cannot make a firm conclusion on the type of CV (i.e., polar versus IP).  We also mention that there is the possibility that this source could be a non-magnetic novalike system, but we consider this possibility less likely as the a majority of hard X-ray detected CVs are mCVs (see e.g., \citealt{2020AdSpR..66.1209D}). Follow-up spectroscopy and photometry can help to better constrain the orbital/spin period of the system to help differentiate between the IP and polar scenario.

\subsection{IGR J20063+3641}

Prior to the analysis performed in this paper, \cite{2018AJ....155..247H} had already identified J20063 as a mCV based on its optical spectrum. The additional follow-up spectra of the source have confirmed that the orbital period of the system is  $P_{\rm orb}=0.731\pm0.015$ d. Furthermore, the X-ray observations have also enabled a detection of the spin period of the WD at $P_{\rm spin}=172.46\pm0.01$ s, which is also detected in the optical observations \citep{2018AJ....155..247H}. Based on the detected orbital and spin periodicities, this system is likely an IP.

The estimated { Gaia} distance to the source,  $d=4.5^{+1.4}_{-1.1}$ kpc  \citep{2020arXiv201205220B}, is a bit larger than the rough upper-limit of $d\approx4$ kpc placed by \cite{2018AJ....155..247H} (which was based on the optical spectrum and the \citealt{2015ApJ...810...25G} reddening map in the direction of the source), but are consistent within errors. Since we cannot place any additional constraints on the distance to the source, we assume a fiducial distance of 4 kpc. At this distance, the source has an observed X-ray luminosity  $L_X\approx3\times10^{34}$ erg s$^{-1}$ in the $0.5-100$ keV energy range, consistent with {\sl INTEGRAL} detected IP luminosities \citep{2020AdSpR..66.1209D}. The IP V2731 Oph has the most similar X-ray luminosity ($L_X=1.2\times10^{34}$ erg s$^{-1}$ in the $0.1-100$ keV energy range; \citealt{2019MNRAS.482.3622S}), orbital period ($P_{\rm orb}=$15.4 hr; \citealt{2005MNRAS.361..141G}), and spin period ($P_{\rm spin}=128$ s; \citealt{2005MNRAS.361..141G}) of any confirmed IPs compared to J20063.  The fact that  J20063 is similar to V2731 Oph may suggest that it has an evolved donor star (see e.g., \citealt{2015ApJ...809...80G,2019ApJ...880..128L}).

\section{Summary}
\label{summary}

Through X-ray observations, we identified J17528 as a new  strong mCV  candidate. { NuSTAR} and { Chandra} X-ray spectra show strong neutral Fe K$\alpha$ emission at 6.4 keV and are well fit by a partially covered bremsstrahlung model, with evidence for a reflection component. The {\sl Chandra} observation has allowed for the optical counterpart of J17528 to be identified and followed-up with MDM optical spectroscopy and photometry. The optical spectrum shows strong H$\alpha$ emission. No orbital or spin periodicity was detected in the X-ray data or in the optical time-series photometry. Assuming a distance of  3 kpc, the source's X-ray luminosity ($L_X\approx10^{34}$ erg s$^{-1}$) is more consistent with those of IPs, but future X-ray and optical observations are needed to confirm this source as an IP by detecting the orbital and WD spin periods.

J20063 was confirmed as a CV system through optical spectroscopy by \cite{2018AJ....155..247H}. The X-ray observations reported here have enabled us to measure the spectrum of this CV, which is well fit by a partially covered bremsstrahlung model and shows evidence of either having an additional blackbody, or, more likely, a reflection component. The X-ray data also allowed for the detection of the WD spin period ($P_{\rm spin}=172.46\pm0.01$ s), while additional MDM optical spectroscopy has allowed for a clear determination of the orbital period at 0.731$\pm0.015$ d. This has allowed us to further classify the source as an IP. Future NIR/IR spectroscopy could be used to further constrain the spectral type of the secondary star  and to place tighter constraints on the source's distance.

 \software{CIAO (v4.11 and the 4.8.3; \citealt{2006SPIE.6270E..1VF}), XSPEC \citep{1996ASPC..101...17A}, Stingray \citep{2019ApJ...881...39H}, NuSTARDAS (v1.8.0)}

\medskip\noindent{\bf Acknowledgments:}
This research has made use of the Palermo BAT Catalogue and database operated at INAF - IASF Palermo.  We thank Nicole Melso and Daniel DeFelippis for obtaining the time-series data on J17528 at MDM Observatory. JH would like to thank Kristin Madsen for useful discussions regarding the different types of stray-light found in { NuSTAR} data and Amy Lien for useful discussions related to the Swift-BAT data. We would like to thank the referee for providing useful comments that  have helped  to improve the overall quality and clarity of the manuscript. JAT and JH acknowledge partial support from the National Aeronautics and Space Administration (NASA) through { Chandra} Award Number GO8-19030X issued by the { Chandra} X-ray Observatory Center, which is operated by the Smithsonian Astrophysical Observatory under NASA contract NAS8-03060. JH acknowledges support from an appointment to the NASA Postdoctoral Program at the Goddard Space Flight Center, administered by the USRA through a contract with NASA.   MC acknowledges financial support from CNES. This work made use of data from the { NuSTAR} mission, a project led by the California Institute of Technology, managed by the Jet Propulsion Laboratory, and funded by the National Aeronautics and Space Administration. We thank the { NuSTAR} Operations, Software and  Calibration teams for support with the execution and analysis of these observations.  This research has made use of the { NuSTAR}  Data Analysis Software (NuSTARDAS) jointly developed by the ASI Science Data Center (ASDC, Italy) and the California Institute of Technology (USA).  This work presents results from the European Space Agency (ESA) space mission Gaia. Gaia data are being processed by the Gaia Data Processing and Analysis Consortium (DPAC). Funding for the DPAC is provided by national institutions, in particular the institutions participating in the Gaia MultiLateral Agreement (MLA). The Gaia mission website is \url{https://www.cosmos.esa.int/gaia}. The Gaia archive website is \url{https://archives.esac.esa.int/gaia}.

\end{document}